\documentstyle[preprint,psfig,aps]{revtex}
\begin{document}

\title{Variational RPA for the Mie resonance in jellium}
\author{G.F. Bertsch,$^{1,4}$ C. Guet,$^2$ and K. Hagino$^{3,4}$}
\address{$^1$
Institute for Nuclear Theory and Department of Physics, \\
University of Washington, Seattle, WA 98195}
\address{$^2$
D\'epartement de Physique Th\'eorique et Appliqu\'ee, CEA-Ile de
France, Bo\^ite Postal 12, 91680 Bruy\`eres le Ch\^atel, France}
\address{$^3$Yukawa Institute for Theoretical Physics, Kyoto
University, Kyoto 606-8502, Japan }
\address{$^4$Institut de Physique Nucl\'eaire, IN2P3-CNRS, \\
Universit\'e Paris-Sud, F-91406 Orsay Cedex, France}

\maketitle

\def\be{\begin{equation}}
\def\ee{\end{equation}}
\def\Do{\Delta\omega}
\def\o0{\omega_0}
\def\o1{\omega_1}
\def\a{\alpha}

\begin{abstract}
The surface plasmon in simple metal clusters is red-shifted from
the Mie frequency, the energy shift being significantly larger
than the usual spill-out correction. Here we develop a variational
approach to the RPA collective
excitations. Using a simple trial form, we obtain analytic
expressions for the energy shift beyond the spill-out
contribution. We find that the additional red shift is
proportional to the spill-out correction and can have the same
order of magnitude.

\end{abstract}

\section{Introduction}

Simple metal clusters exhibit a strong peak in their optical
response that corresponds to a collective oscillation of the
valence electrons with respect to a neutralizing positively
charged background.  Classically, the frequency of the
oscillation is given by the Mie resonance formula\cite{Kr95,Be94},
\be
\omega_{\rm Mie}^{2}={4\pi ne^{2}\over 3m}
\ee
where $n$ is the density of a
homogeneous electron gas. Quantum finite size effects lead to a
red shift of this frequency as well as to a redistribution of
the oscillator strength ($f$) into closely lying dipole states.
Moments of the oscillator strength distribution $M_k = \sum_i
\omega_i^{k-1} f_i$ provide useful information. The first moment
$M_{1}$, which measures the integral of the $f$-distribution,
equals the number of electrons (Thomas-Reiche-Kuhn sum rule).
The mean square frequency $\langle\omega^{2}\rangle=M_{3}/M_{1}$ is given by
the overlap integral of the positive ionic charge distribution
and the exact ground state electronic density\cite{Be94}. Within
an ionic background approximated by a jellium sphere, the mean
square frequency is thus exactly related to the square Mie
frequency by

\begin{equation}
\langle\omega^{2}\rangle=\omega_{\rm Mie}^{2}\left(  1-\frac{\Delta N}{N}\right)
\end{equation}
where $\Delta N/{N}$ is the fraction of electrons in the ground
state that is outside the jellium sphere radius.
We called the corresponding energy shift $\Delta \omega_{so}$
(``spill-out"):
\be
\Delta \omega_{so} = \omega_{\rm Mie}(1-\sqrt{1-\Delta N/N})
\ee

The actual red shifts are considerably larger than this.
For sake of illustrating the discussion let us consider
the sodium cluster Na$^+_{21}$ for which detailed photoabsorption data is
available\cite{Kn89,Se91,Br89,Sc99,Re95}.  The Mie frequency is at 3.5 eV,
taking the density corresponding to $r_s=3.93$ a.u., while the measured
resonance is a peak 2.65 eV having width of about 0.3 eV (FWHM). Thus
there is a red shift of 24\%, which may be compared with a 9\%
red shift predicted by eq. (3) using jellium wave functions.
 To a large extent clusters with a ``magic''
number of valence electrons behave optically as close shell
spherical jellium spheres. The experimental photoabsorption
spectra for these clusters are well described within the linear
response theory using either the time-dependent local-density
approximation (TDLDA)\cite{Ek85,Yan91,Yab96} or the random phase
approximation with exact exchange (RPAE)\cite{Gu92,Ma95}. Red
shifts of 14\% and 18\% are predicted by time-dependent density
functional theory \cite{be90} and by the random phase
approximation \cite{Gu92,Ma95}, respectively.
The oscillator strength distributions in the RPA calculations are
typically dominated by a few close states that exhaust almost all
of the sum rule. It is this concentration of strength, which we
can identify as a dipole surface plasmon, that will be of interest
in this paper. It is worth of note that singling out a collective
state is not always possible even in small clusters. Whenever the
collective state lies within a region of high level density, there
is a strong fragmentation into p-h states (Landau damping) and
several excited states may share evenly the strength. We will deal
with this problem of the definition of the collective state later
by proposing a model in which there is no particle-hole
fragmentation.
\newline

Anharmonic effects in metallic clusters
were studied recently by Gerchikov, {\it et al.}, \cite{ge02} 
making use of a coordinate transformation to separate
center of mass (c.m.) and intrinsic motion.  The authors show that in
absence of coupling between c.m. motion and intrinsic excitations
the surface plasmon associated with a jellium sphere has a single
peak which is red-shifted with respect to the Mie frequency by the
spill-out electrons, Eq. (3). Turning on the coupling yields a
further red shift which indeed is larger in magnitude than the
spill-out contribution. Concomitantly, there is a partial transfer
of strength into states of higher energy preserving the sum rule,
Eq. (2). The approach requires the spectrum of 
excitations in the intrinsic coordinates, which  were obtained
by projection on the computed wave functions of the numerical RPAE.

Another interesting approach to the coupling between the collective
and noncollective degrees of freedom was developed by Kurasawa,
{\it et al.}, \cite{ku97}, following the Tomonaga expansion of
the Hamiltonian.  The collective coordinate is taken as the cm
coordinate, as in ref. \cite{ge02}, and the coefficients of the
harmonic terms in the Hamiltonian  yield Eq. (2) for the frequency.
The authors derive expressions
for the coupling terms in the Hamiltonian and use them to
estimate
the variance of the  Hamiltonian in the collective state.
They find that the variance decreases with size of the cluster
as $1/R$, where $R$ is
the radius of the ion distribution.  Both the width of the Mie
and its shift are obviously related to the variance of H,
but further assumptions are needed to make a quantitative connection.

In the present paper we wish to find an
analytic estimate of the red shift, keep as far as possible the
ordinary formulation of RPA, and not singling out a collective
state in the Hamiltonian.  Our approach will be 
a variational RPA theory, which we present in the next section.
The rest of the paper is organized as follows.  
In Section III we apply the formalism to a system of
interacting electrons. The model Hamiltonian describes interacting
electrons confined in a pure harmonic potential, whereas the
perturbation corrects for the jellium confinement. The model RPA
solution is derived analytically and first and second order
corrections of the frequency shift are given.

\section{Variational RPA }
In this section, we establish our notation for the RPA theory of
excitations and develop a variational expression for perturbations to
the collective excitation frequency.  The perturbation behaves somewhat
differently in RPA
than in conventional matrix Hamiltonians because the RPA operator
is not Hermitean.

As usual, the starting point is a mean field theory whose ground
state is represented by an orbital set ${\phi_i}$ satisfying the
orbital equations \be h[\rho_0] \phi_i = \epsilon_i \phi_i \ee
where $\rho_0 = \sum_i |\phi_i(r)|^2$.  The RPA equations are
obtained by considering small deviations from the ground state,
\be \phi_i \rightarrow \phi_i + \lambda (x_i e^{-i\omega t} +y_i
e^{i \omega t}). \ee Here $x_i,y_i$ are vectors in whatever space
($r$-space,orbital occupation number,...) is used to represent
$\phi_i$.  The RPA equations can be expressed as \be
(h[\rho_0]-\epsilon_i) x_i + \delta \rho* {\delta h\over \delta
\rho} * \phi_i =\omega x_i \ee
$$
-(h[\rho_0]-\epsilon_i)y_i -\delta \rho* {\delta h\over \delta \rho} *
\phi_i =\omega y_i
$$
where the transition density $\delta \rho $ is defined by
$$
\delta \rho = \sum_i \phi_i (x_i + y_i)
$$
and the symbol $*$ denotes an operator or matrix multiplication.
Eq. (6) represents
linear eigenvalue problem for a nonhermitean operator $R$ and
the vector $|z\rangle=(x_1,y_1,x_2,y_2,...)$.
We will write the equations compactly as
$$
R |z\rangle = \omega |z\rangle.
$$
For a nonhermitean operator, the adjoint vector $\langle z | $ is
defined as the eigenvector of the adjoint equation, $ \langle z |R
= \omega \langle z|.$ From the symmetry of $R$ it is easy to see
that it is given by $\langle z | =
(x_1,-y_1,x_2,-y_2,...)^\dagger$.

We now ask how to construct a perturbation theory starting from
the zero-order wave function $|z_0 \rangle$  that is the solution
of an unperturbed $R_0$ with eigenfrequency  $\omega_0$. If we had
the complete spectrum of $R_0$, the perturbation series for
$R=R_0+\Delta R$ could be written down in the usual way,
$$
| z\rangle = |z_0 \rangle + \sum_\alpha | z_\alpha\rangle {\langle z_\alpha
| \Delta R | z \rangle \over \omega_0 - \omega_\alpha },
$$
etc. This is in fact what is done in ref. \cite{ge02}.
However, this requires diagonalizing $R_0$ which in general can
only be done numerically.

Instead we shall estimate the energy perturbation using a
variational expression for the frequency,
\be
\omega = \min_w {\langle z_0 + \lambda w| R | z_0 + \lambda w
\rangle \over \langle
z_0+ \lambda w
| z_0 + \lambda w \rangle },
\ee
where $| w \rangle$ is a vector to be specified later and $\lambda$
is to be varied to minimize the expression.  Carrying out the variation
and assuming that the perturbation is
small, the value of $\lambda$ at the minimum is given by
\be
\lambda \approx -{\langle z_0 |R |w\rangle -\omega_0 \langle z_0 |w\rangle
\over  \langle w | R
w \rangle -\o1 \langle w| w\rangle }
\ee
and the energy shift is
\be
\omega \approx \omega_0 +\langle z_0 | \Delta R | z_0 \rangle -
 {(\langle z_0 |R w\rangle -\o1 \langle z_0 |w\rangle )^2
\over  \langle w | R
w \rangle -\o1 \langle w | w \rangle} .
\ee
Here, $\omega_1\equiv \langle z_0|R|z_0 \rangle = \omega_0 +
\langle z_0|\Delta R|z_0 \rangle.$

The next question is how to choose the perturbation $| w\rangle$.
With ordinary Hamiltonians, one can construct a two-state
perturbation theory using the vector obtained by applying $\Delta
R$ to the unperturbed vector, $|w\rangle = \Delta R|z_0\rangle$.
However, we will see in the next section that this fails
completely for the RPA operator.  Instead, we will find that an
approximation that gives qualitatively acceptable results can be
made by taking only the $x$-component of the vector defined by
applying $\Delta R$ to $|z_0\rangle$.

\section{Collective limit of the surface plasmon}

We apply the RPA variational perturbation theory derived in the
previous section to the surface plasmon of small metal clusters.
We write the single particle Hamiltonian as
\begin{eqnarray}
h&=&h_0 + \Delta V(r), \\
h_0&=&-\frac{\hbar^2}{2m}\nabla^2 + \frac{1}{2}m\omega_0^2r^2 +v *\rho_0,
\label{h0}
\end{eqnarray}
where $v *\rho_0$ is the mean field potential,
\begin{equation}
v *\rho_0=\int v(r,r')\rho_0(r') \,d^3r'.
\end{equation}
Here $v$ is the electron-electron interaction, which may contain
an exchange-correlation contribution from density functional
theory. In this paper, we throughout use the jellium model for the
ionic background, and also assume that the ion and the electron
densities are both spherical. $\omega_0$ and $\Delta V(r)$ are
then given by $\omega_0=Ze^2/mR^3$ and
\begin{equation}
\Delta V(r) = \left[-\frac{Ze^2}{r}-\left(-\frac{3}{2}+\frac{r^2}{2R^2}
\right)\frac{Ze^2}{R}\right]\,\theta(r-R),
\label{deltaV}
\end{equation}
respectively, $R$ being the sharp-cutoff radius for the ion distribution.

The RPA equations can be solved exactly for the Mie resonance if
$h$ is replaced by $h_0$.
The solution is
\begin{equation}
|z_0\rangle \equiv \left(\matrix{ x \cr y \cr} \right) =
-\sqrt{\frac{m\omega_0}{2N}} \left(\matrix{ z\phi \cr -z\phi \cr}
\right) + \sqrt{\frac{1}{2Nm\omega_0}} \left(\matrix{
\partial_z\phi \cr \partial_z\phi \cr} \right),
\label{collective}
\end{equation}
associated with the eigenfrequency $\omega_0$. Notice that the
eigenfrequency $\omega_0$ is the same as the harmonic oscillation
frequency in Eq. (\ref{h0}), agreeing with the Kohn's theorem
\cite{K61,D94,V95,vi95,vi01}.

To prove that the collective solution (\ref{collective}) satisfies
the RPA equation, we use the following identity which results from
the Hartree-Fock equation,
\begin{equation}
(h-\epsilon)(\hat{A}\phi) = [h,\hat{A}]\phi. \label{apsi}
\end{equation}
Here $\hat{A}$ is any one body operator. This yields
\begin{eqnarray}
(h_0-\epsilon)(z\phi) &=& -\frac{1}{m}\partial_z\phi, \\
(h_0-\epsilon)(\partial_z\phi) &=&
-(m\omega_0^2z+(v * \partial_z\rho_0))\phi.
\label{collective2}
\end{eqnarray}
In the last step, we used the fact that the interaction $v$ is
translationally invariant. Notice that the transition density is
proportional to $\partial_z\rho_0$ for the collective solution
(\ref{collective}). The second term in Eq. (\ref{collective2}) is
thus exactly canceled by the residual interaction term in the RPA
equations, proving that the collective ansatz (\ref{collective})
is indeed the eigenfunction of the RPA matrix $R_0$ with the
eigenvalue $\omega_0$.

The familiar formula relating the red-shift to the electron
spill-out probability can be recovered from the expectation value
of the original RPA matrix,
\begin{equation}
\frac{(z_0|Rz_0)}{(z_0|z_0)}=\omega_0+\Delta\omega.
\label{expectation}
\end{equation}
However, the wave function $z_0$ must be taken with the collective
ansatz applied to the Hamiltonian $h$. This is different from the
$z_0$ defined in Eq. (\ref{collective}), which was based on the
Hamiltonian $h_0$. In the following, we have no further use for
the original $z_0$ and we will use the same name here. Applying
the RPA operator $R$ to $z_0$, we find
\begin{equation}
R|z_0\rangle = \omega_0|z_0\rangle + |u\rangle,
\label{Rz0}
\end{equation}
where $u$ is given by
\begin{equation}
|u\rangle = -
\sqrt{\frac{1}{2Nm\omega_0}}\,\frac{1}{r}\frac{d\Delta V}{dr}
\left(\matrix{ z\phi \cr -z\phi \cr} \right).
\label{u-perturbation}
\end{equation}
The expectation value eq. (\ref{expectation}) then reduces to
\begin{equation}
\Delta\omega = (z_0|u) = -\omega_0\,\frac{\Delta N}{2N},
\label{spillout}
\end{equation}
with
\begin{equation}
\Delta N = \int^{\infty}_R4\pi r^2dr \,\rho_0(r).
\end{equation}
%
Eq. (\ref{spillout}) is just the well-known spill-out formula,
Eq.(3), to the first order in $\Delta N/N$.

\section{Evaluation of the integrals}

We now consider the frequency shift in the second order
perturbation. Obvious possibilities for the perturbation are
$w_0\equiv(y,x)$ and $u$, but we find that neither produces a
significant energy shift. The problem with $u$ is that the $x$
component is tied to the $y$ component in Eq.
(\ref{u-perturbation}). In fact, the energetics are such the $y$
perturbation is much less than the $x$ perturbation. In order to
avoid this undesirable feature, as we mentioned in Sec. II, we
simply take the $x$ component of $u$ for the perturbation. That
is, we use
\begin{equation}
|\tilde{u}\rangle \equiv
\frac{1}{r}\frac{d\Delta V}{dr}
\left(\matrix{
z\phi \cr 0 \cr} \right)
=
\frac{\partial\Delta V}{\partial z}
\left(\matrix{
\phi \cr 0 \cr} \right),
\end{equation}
for the $|w\rangle$ in the variational formula (7). With this
perturbed wave function, after performing the angular integration,
we find the three integrals in the formula to be
\begin{eqnarray}
\langle z_0|\tilde{u}\rangle
&=&
-\sqrt{\frac{m\omega_0}{2N}}\frac{4\pi}{3}
\int^{\infty}_Rr^3dr\,\frac{d\Delta V}{dr}\,\rho_0(r) \nonumber \\
&& +
\sqrt{\frac{1}{2Nm\omega_0}}
\frac{2\pi}{3}
\int^{\infty}_Rr^2dr\,\frac{d\Delta V}{dr}\frac{d\rho_0}{dr},
\label{z0u} \\
\langle \tilde{u}|\tilde{u}\rangle
&=&\frac{4\pi}{3}
\int^{\infty}_Rr^2dr\,\left(\frac{d\Delta V}{dr}\right)^2\rho_0(r),
\label{uu} \\
\langle z_0|R\tilde{u}\rangle
&=&
\omega_0\langle z_0|\tilde{u}\rangle
+\langle u|\tilde{u}\rangle
=
\omega_0\langle z_0|\tilde{u}\rangle
-\sqrt{\frac{1}{2Nm\omega_0}}\,
\langle \tilde{u}|\tilde{u}\rangle.
\label{z0Ru}
\end{eqnarray}
In deriving Eq.(\ref{z0Ru}), we have used Eq.(\ref{Rz0}).
We also need to compute $\langle \tilde{u}|R|\tilde{u}\rangle$ in
order to estimate the energy shift. Neglecting the residual
interaction in the RPA operator $R$, this is expressed as
\begin{equation}
\langle \tilde{u}|R|\tilde{u}\rangle\sim
\langle
\frac{\partial\Delta V}{\partial z}\phi|
h-\epsilon|
\frac{\partial \Delta V}{\partial z}\phi\rangle.
\end{equation}
We use Eq.(\ref{apsi}) to evaluate the action of the Hamiltonian
$h$ onto the $\tilde{u}$. This yields
\begin{equation}
(h-\epsilon)\,|\frac{\partial \Delta V}{dz}\phi\rangle
=-\frac{1}{2m}\left[\left(\nabla^2\frac{\partial \Delta V}
{\partial z}\right)
+2\left(\nabla \frac{\partial \Delta V}{\partial z}\right)\cdot
\nabla\right]\,|\phi\rangle.
\end{equation}
Notice that the first term vanishes for the jellium model
(\ref{deltaV}). We thus finally have
\begin{eqnarray}
\langle \tilde{u}|R|\tilde{u}\rangle
&=&
-\frac{1}{2m}\int d^3r\,\frac{\partial \Delta V}{\partial z}
\left(\nabla \frac{\partial \Delta V}{\partial z}\right)\cdot
\nabla \rho_0, \\
&=&
-\frac{1}{2m}\frac{4\pi}{3}\int^\infty_Rr^2dr\,
\frac{d\Delta V}{dr}\frac{d^2\Delta V}{dr^2}\frac{d\rho_0}{dr}.
\label{uRu}
\end{eqnarray}

In order to get a simple analytic formula for the energy shift,
we estimate Eqs. (\ref{z0u}), (\ref{uu}), (\ref{z0Ru}), and
(\ref{uRu}) assuming that the density $\rho_0$ in the surface
region is given by
\begin{equation}
\rho_0(r)\sim A e^{-2\kappa (r-R)}~~~~~~~~(r \ge R),
\label{density0}
\end{equation}
with $\kappa^2/2m = \epsilon$, where $\epsilon$ is the ionization
energy. In order to simplify the algebra, we also expand $\Delta V$
and take the first term,
\begin{equation}
\frac{d\Delta V}{dr}\sim -3m\omega_0^2(r-R).
\end{equation}
These approximations lead to the following analytic expressions,
\begin{eqnarray}
\langle z_0|\tilde{u}\rangle
&=&
4\pi A m\omega_0^2\, \sqrt{\frac{m\omega_0}{2N}}
\left\{\frac{R^3}{4\kappa^2}
-\frac{3R^2}{4\kappa^3}+\frac{9R}{8\kappa^4}+\frac{3}{4\kappa^5}\right.
\nonumber \\
&&\left.+2\cdot\frac{\epsilon}{\omega_0}\left(
\frac{R^2}{4\kappa^3}+\frac{R}{2\kappa^4}+\frac{3}{8\kappa^5}\right)\right\}
,
\\
\langle \tilde{u}|\tilde{u}\rangle
&=&
12\pi A m^2\omega_0^4\left(
\frac{R^2}{4\kappa^3}+\frac{3R}{4\kappa^4}+\frac{3}{4\kappa^5}\right),
\\
\langle \tilde{u}|R|\tilde{u}\rangle
&=&
12\pi A m\omega_0^4
\left(
\frac{R^2}{4\kappa}+\frac{R}{2\kappa^2}+\frac{3}{8\kappa^3}\right).
\end{eqnarray}
Note that with the density (\ref{density0})
the spill-out electron number $\Delta N$ is given by
\begin{equation}
\Delta N=4\pi A
\left(
\frac{R^2}{2\kappa}+\frac{R}{2\kappa^2}+\frac{1}{4\kappa^3}\right).
\end{equation}
Retaining only the leading order of $1/\kappa R$, we thus have
\begin{eqnarray}
\langle z_0|\tilde{u}\rangle
&=&
m\omega_0^2\, \sqrt{\frac{m\omega_0}{2N}}
\frac{R}{2\kappa}\,\Delta N, \\
\langle \tilde{u}|\tilde{u}\rangle
&=&
3 m^2\omega_0^4\,\frac{\Delta N}{2\kappa^2}, \\
\langle \tilde{u}|u\rangle
&=&
-3m\omega_0^3\, \sqrt{\frac{m\omega_0}{2N}}
\frac{\Delta N}{2\kappa^2}, \\
\langle \tilde{u}|R|\tilde{u}\rangle
&=&
\frac{3}{2}m\omega_0^4\Delta N.
\end{eqnarray}
Substituting these expressions into Eq. (9), we finally obtain
\begin{equation}
\omega = \omega_1-\frac{3}{16-8\cdot\omega_0/\epsilon}
\left(\frac{\omega_0}{\epsilon}\right)^2\cdot \omega_0
\frac{\Delta N}{N}.
\end{equation}
This is our main result. Note that the perturbation theory breaks
down at $\epsilon=\omega_0/2$. In realistic situations discussed
in the next section, $\epsilon$ is always close to $\omega_0$, and
the perturbation theory should work in principle.

\section{Numerical comparison with the RPA solutions}

To assess the reliability of the variational shifts, we have
numerically solved the RPA equations for the jellium model, using
the computer program JellyRpa \cite{be90}.  A typical spectrum is
shown in Fig. 1. This represents Na$_{20}$ as a system of 20
electrons in a background spherical charge distribution with a
density corresponding to $r_s=3.93$ a.u. and total charge $Q=20$.
The strength function includes an artificial width of $\Gamma=0.1$
eV for display purposes.  The Mie frequency, Eq. (1), is indicated
by $\omega_0$, while the prediction of the spill-out formula,
Eq.(3), is shown as $\omega_{so}$ in the figure. One sees that the
strength function is fragmented into two large components that are
considerably red-shifted from the Mie frequency, and smaller
contributions at higher frequencies. The corresponding spectrum
with the jellium background potential replaced by a pure harmonic
potential is shown by the dashed line.  The numerical RPA
frequency agrees very well with the Mie value in this case,
showing that the numerical algorithms used in JellyRpa are
sufficiently accurate for our purposes. The red shift can be more
easily displayed by a plot of the integrated strength function,
shown in the lower panel of the figure. If we define the shift as
the point where the integrated strength reaches half of the
maximum value, it corresponds to $\delta \omega = 0.166\,
\omega_{\rm Mie}$.  On the other hand, the collective formula for
the red shift, Eq. (3), only gives $\delta \omega =
0.058\,\omega_{\rm Mie}$, when the integral for $\Delta N$ is
evaluated with the ground state density.

The strength becomes increasingly fragmented in heavier
clusters, making a precise definition of the red shift
problematic.  We therefore have simplified the jellium model
in our numerical computations to see the effects of the
shift  without the fragmentation of the
strength that occurs physically.  To this end we put all the electrons
in the lowest $s$-orbital, treating them as bosons.  Otherwise,
the model is the same as the usual jellium model, with the
electron orbitals determined self-consistently in a background charge
density of a uniform sphere.  This model is easily implemented with
JellyRpa by assigning the occupation probabilities of
the orbitals appropriately.
Taking the density parameter as $r_s=3.93$ a.u.,
appropriate for sodium clusters, one finds that the ionization potential
is rather close to the value of the usual (fermionic) jellium model.
For example, in the cluster with $N=20$ atoms, the ionization potential
$\epsilon$ has a value 2.84 eV for usual jellium model and the value
4.11 eV for our simplified $s$-wave treatment.

The results of the numerical calculation with the full effect of the
surface are shown in Fig. 2 as the solid line.
The collective spill-out
correction from Eq. (3) is also shown as the dotted line.  One sees that
the additional shift due to the wave function perturbation is
comparable to the spill-out correction, and has a similar
$N$-dependence.  The shift given by the variational formula Eq.
(9) is shown by the dashed line.  The functional dependence
predicted by the formula is confirmed by the numerical
calculations, but the coefficient of $N$ is too small by a factor
of two or so.

\section{Concluding remarks}

We have developed a variational approach to treat perturbations
to the collective RPA wave functions, and have applied it to the
surface plasmon in small metal clusters. Our zeroth order solution
is the same as that used by Gerchikov {\it et al.} \cite{ge02} and 
Kurasawa {\it et al.} \cite{ku97}.  
It corresponds to the center
of mass motion, and is the exact RPA solution when the ionic
background potential is a harmonic oscillator. The deviation of
the background potential from the harmonic shape is responsible
for the perturbation. The first order perturbation yields the
well-known spill-out formula for the plasmon frequency, as was
also shown in Refs. \cite{ge02,ku97}. The higher order corrections
lead to the additional energy shift of the frequency \cite{ge02},
the anharmonicity of the spectrum \cite{ge02}, and the
fragmentation of the strength \cite{ku97}. Those effects were
studied in Refs. \cite{ge02,ku97} by considering explicitly the
couplings between the center of mass and the intrinsic motions. In
this paper, we assumed some analytic form for the perturbation and
determined its coefficient variationally. We found that this
approach qualitatively accounts for the red shift of the
collective frequency, but its magnitude came out too small by
about a factor of two.

In order to have a more quantitative result, one would have to
improve the variational wave function. An obvious way is to
introduce more than one term. Our method may be viewed as the
first iteration of any iterative method for RPA
\cite{JBH99,MIHY02,IH03}. One may need more than one iteration to
get a convergence and thus a sufficiently large energy shift.
Another possible way is to construct the perturbed wave function
based on the local RPA. The authors of Ref. \cite{re90} expanded
the collective operator with local functions and solved a secular
equation to determine the frequency. They showed that the
expansion of the collective operator with three functions,
$r\cos\theta, r^3\cos\theta$, and $r^5\cos\theta$, gives a
satisfactory result for the collective frequency.

The method developed in this paper is general, and is not
restricted to the surface plasmon in micro clusters. One
interesting application may be to the giant dipole resonance in
atomic nuclei. In heavy nuclei, the mass dependence of the
isovector dipole frequency deviates from the prediction of the
Goldhaber-Teller model, that is based on a simple c.m.
motion\cite{BS76,MS77}. The shift of collective frequency can be
attributed to the effect of deviation of the mean-field potential
from the harmonic oscillator, and a similar treatment as the
present one is possible.

\section*{Acknowledgments}

We would like to acknowledge discussions with Nguyen Van Giai, N.
Vinh Mau, P. Schuck, and M. Grasso. K.H. thanks the IPN Orsay for
their warm hospitality and financial support. G.F.B. also thanks
the IPN Orsay as well as CEA Ile de France for their hospitality
and financial support. Additional financial support from the
Guggenheim Foundation and the U.S. Department of Energy (G.F.B.)
and from the the Kyoto University Foundation (K.H.) is
acknowledged.

\begin{figure}
  \begin{center}
    \leavevmode
    \parbox{0.9\textwidth}
           {\psfig{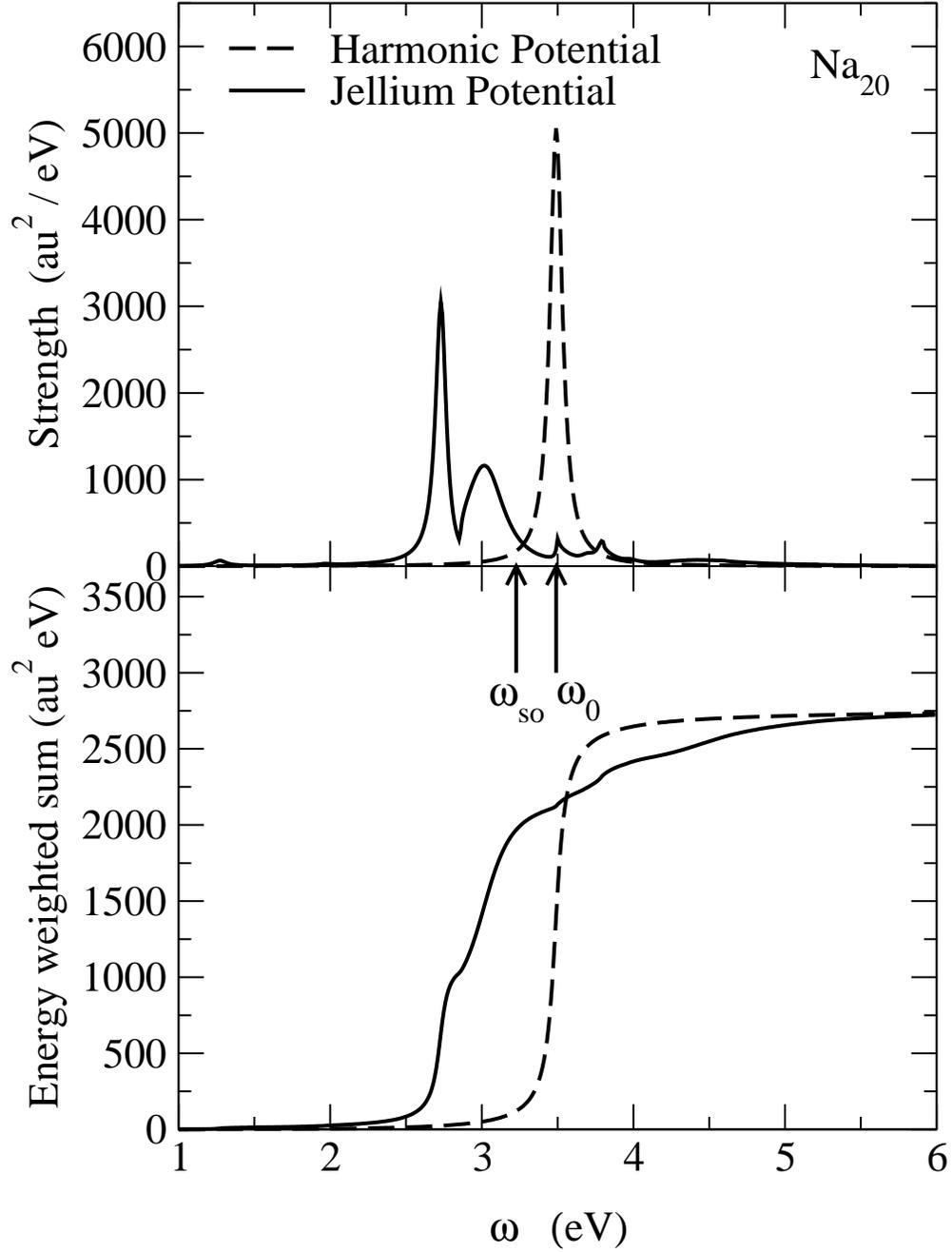}}
  \end{center}
\caption{Strength function of Na$_{20}$ in the jellium model.
Upper panel shows the dipole strength function, broadened by a
artifical width.  Lower panel shows the integerated strength
function.  Dashed line is the results of the computation
in which the jellium background potential is replaced by
a harmonic oscillator.
}
\label{jellium}
\end{figure}

\begin{figure}
  \begin{center}
    \leavevmode
    \parbox{0.9\textwidth}
           {\psfig{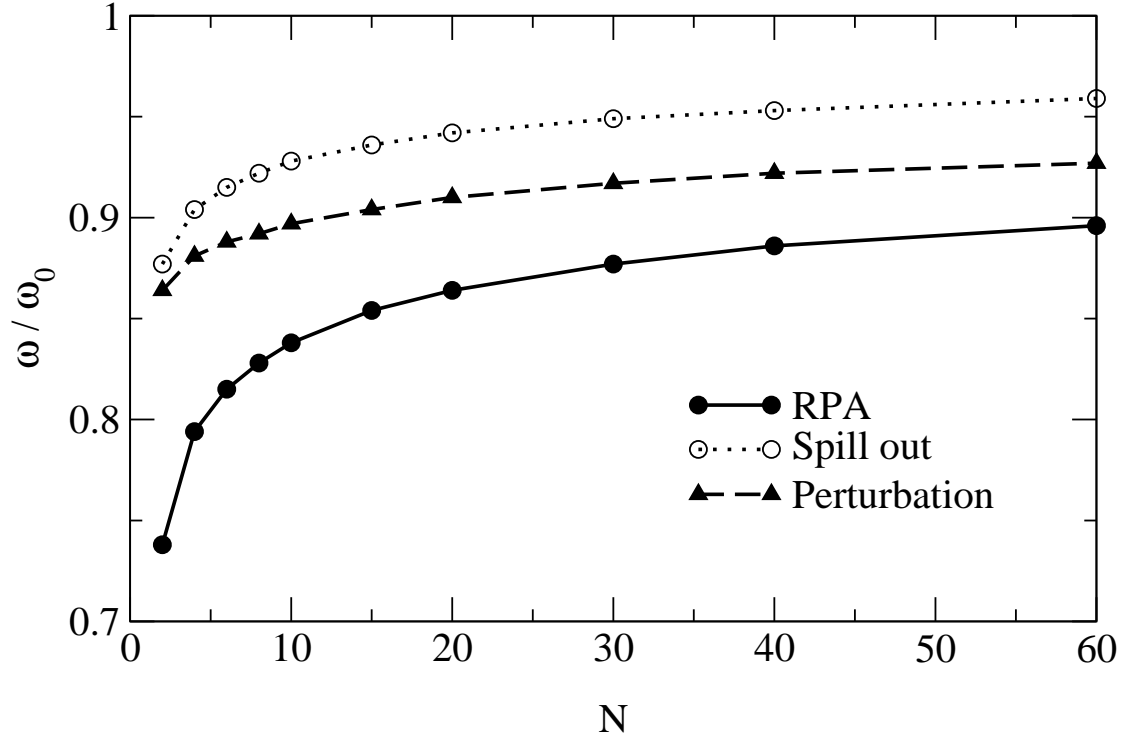}}
  \end{center}
\caption{Collective excitation frequency in the $s$-wave jellium
model as a function of $N$.  The solid line is the result of the
numerical calculation.  This is compared with the spill-out
formula eq. (3) and the perturbation formula eq. (9) as the
dotted
and dashed lines, respectively.
}
\label{systematics}
\end{figure}

\end{document}